\documentclass[%
 prx,
 amsmath,amssymb,
 reprint,%
 superscriptaddress,
]{revtex4-2}

\usepackage{graphicx}
\usepackage{dcolumn}
\usepackage{bm}
\usepackage{tabularx}
\usepackage{graphicx}
\usepackage{multirow}
\usepackage{siunitx,color} 
\usepackage{mathrsfs}
\usepackage[colorlinks=true,citecolor=blue,linkcolor=blue
]{hyperref}

\begin{document}
\bibliographystyle{apsrev4-2.bst}

\title{Half-Valley Ohmic Contact and\\ Contact-Limited Valley-Contrasting Current Injection}

\author{Xukun Feng}
\affiliation{Science, Mathematics and Technology Cluster, Singapore University of Technology and Design, Singapore 487372.}

\author{Chit Siong Lau}
\affiliation{Institute of Materials Research and Engineering, Agency for Science, Technology and Research (A*STAR), Singapore 138634.}

\author{Shi-Jun Liang}
\affiliation{National Laboratory of Solid State Microstructures, School of Physics, Collaborative Innovation Center of Advanced Microstructures, Nanjing University, Nanjing 210093, China.}

\author{Ching Hua Lee}
\email{phylch@nus.edu.sg}
\affiliation{School of Physical Science, National University of Singapore, Singapore 117551.}

\author{Shengyuan A. Yang}
\email{shengyuan\_yang@sutd.edu.sg}
\affiliation{Science, Mathematics and Technology Cluster, Singapore University of Technology and Design, Singapore 487372.}

\author{Yee Sin Ang}
\email{yeesin\_ang@sutd.edu.sg}
\affiliation{Science, Mathematics and Technology Cluster, Singapore University of Technology and Design, Singapore 487372.}

\begin{abstract}

Two-dimensional (2D) ferrovalley semiconductor (FVSC) with spontaneous valley polarization offers an exciting material platform for probing Berry phase physics. How FVSC can be incorporated in valleytronic device applications, however, remain an open question. Here we generalize the concept of metal/semiconductor (MS) contact into the realm of valleytronics. We propose a \emph{half-valley Ohmic contact} based on FVSC/graphene heterostructure where the two valleys of FVSC \emph{separately} forms Ohmic and Schottky contacts with those of graphene, thus allowing current to be \emph{valley-selectively} injected through the `Ohmic' valley while being blocked in the `Schottky' valley. We develop a theory of \emph{contact-limited valley-contrasting current injection} and demonstrate that such transport mechanism can produce gate-tunable valley-polarized injection current. Using RuCl$_2$/graphene heterostructure as an example, we illustrate a device concept of \emph{valleytronic barristor} where high valley polarization efficiency and sizable current on/off ratio, can be achieved under experimentally feasible electrostatic gating conditions. These findings uncover contact-limited valley-contrasting current injection as an efficient mechanism for valley polarization manipulation, and reveals the potential of valleytronic MS contact as a functional building block of valleytronic device technology.  

\end{abstract}
\maketitle

\section{Introduction}

Valleytronics is an emerging device concept that harnesses the valley degree of freedom for information encoding and processing \cite{xiao2007valley,yao2008valley,xiao2012coupled,Liu2015,schaibley2016valleytronics,li2013coupling,rycerz2007valley,Li2020Room,ang2017valleytronics}. Valley-contrasting phenomena, such as anomalous valley hall effect (AVHE) \cite{Mak2014}, has been studied extensively in two-dimensional (2D) materials \cite{mak2010atomically,Wang2012,goh2023valleytronics}, and can be generated via a plethora of structural \cite{rycerz2007valley,Zhu2012, Ju2015, Yu2020, zhang20222d}, electrical \cite{ang2017valleytronics,tao2019two}, mechanical \cite{Ghaemi2012, doi:10.1063/1.3473725}, magnetic \cite{Cheng2014,Zhang2016,Peng2018, li2018valley}, and optical \cite{Zeng2012, bussolotti2023band} mechanisms. Recent discovery of ferrovalley semiconductor (FVSC) \cite{Tong2016} unveils an exciting platform to study valley-constrating phenomena and Berry phase physics in the 2D \emph{flatland}. Due to the simultaneous breaking of spatial inversion and time reversal symmetries, the $K$ and $K'$ valley degeneracy is broken in FVSC, and valley polarization emerges \emph{spontaneously} without any external regulating mechanisms \cite{Zhang2019,Luo2020,Hu2020,Zhao2019,Peng2020Intrinsic,Cheng2021,Zang2021,Wang2020,sheng2022strain,li2022robust,he2021two, feng2022valley}. 

The spontaneous valley polarization in FVSC \cite{MacNeill2015,Li2020Electrical} is particularly much sought after for producing valley polarized current (VPC) -- an electrical current with carriers dominantly from only one valley.
Despite tremendous efforts in the search for FVSC candidates and the high anticipation of FVSC as a building block of valleytronics, the questions of (i) how FVSC forms contact with metal electrodes -- an omnipresent component of modern electronics; (ii) how contact-limited charge injection occur \cite{ang2018universal,ang2021physics}; and (iii) how FVSC/metal contact can be incorporated in the design of valleytronic devices  remain largely open thus far \cite{bussolotti2018roadmap}.

\begin{figure*}
\includegraphics[width=0.95\textwidth]{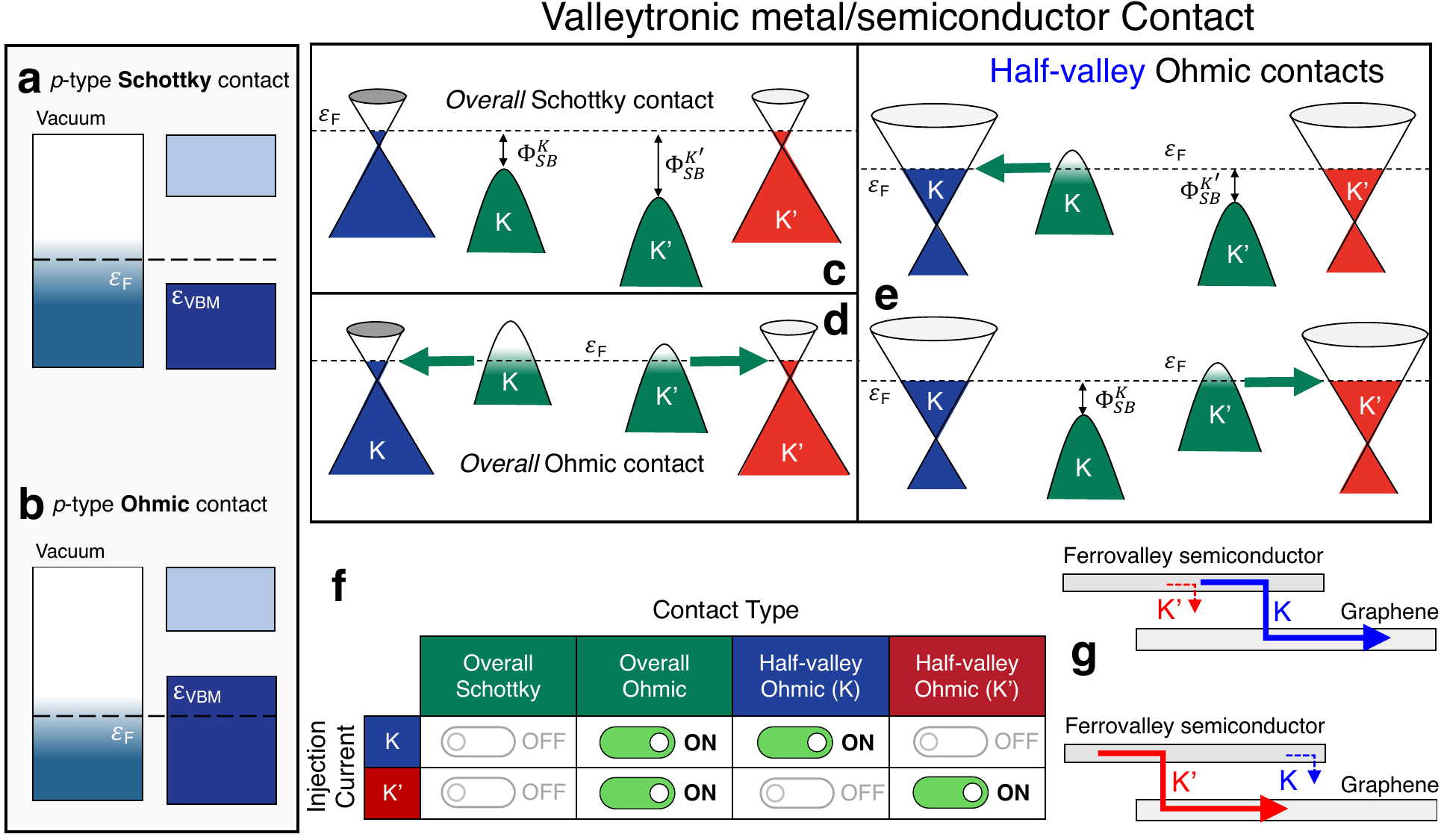}
\caption{\label{fig:1} \textbf{Concept of half-valley Ohmic contact.} Schematic diagrams of \textbf{a} \textit{p}-type Schottky contact and \textbf{b} \textit{p}-type Ohmic contact. \textbf{c}-\textbf{f} Schematic diagrams of overall Schottky contact, overall Ohmic contact and half-valley Ohmic contact. \textbf{g} The `on' and `off' states of current injection in $K$ and $K'$ valleys regarding to three contact types. \textbf{h} and \textbf{i} Valley-selective current injection for $K$ and $K$' valleys. }
\end{figure*}

In this work, we addressed the above questions by generalizing the conventional MS contact into the realm of valleytronics. We show that the synergy of FVSC and metal contact can achieve efficient valley degree of freedom manipulation.
We propose the concept of \emph{half-valley Ohmic contact} in which the $K$ and $K'$ valleys of FVSC separately forms Schottky and Ohmic contacts with a contacting metal of matching $K$ and $K'$ valleys. 
Electrical current can be injected through the `Ohmic valley' while being suppressed at the `Schottky valley'.
We develop a model of contact-limited valley-contrasting current injection across graphene/FVSC, and show that such process provides a mechanism to manipulate the valley polarization of the injection current.
Using graphene/RuCl$_2$ contact heterostructures as a model example, the material realization of half-valley Ohmic contact is demonstrated.
We propose a device concept of \emph{valleytronic barristor} (i.e. `barrier' + `transistor' \cite{gr_bar}) in which the injection current can be switched on and off, and the VPC can be tuned all-electrically under experimentally feasible electrostatic gating conditions. 
Our findings unveil a wealth of MS contact species, notably the half-valley Ohmic contact, which emerges when generalizing the concept of MS contact into the valleytronic realm.
The half-valley Ohmic contact proposed in this work shall provide a useful functional building block of valleytronic devices, and shall serve as a harbinger for a new class of valleytronic devices that manipulates valley degree of freedom via contact-limited valley-contrasting charge injection process.

\begin{figure*}
\includegraphics[width=0.858\textwidth]{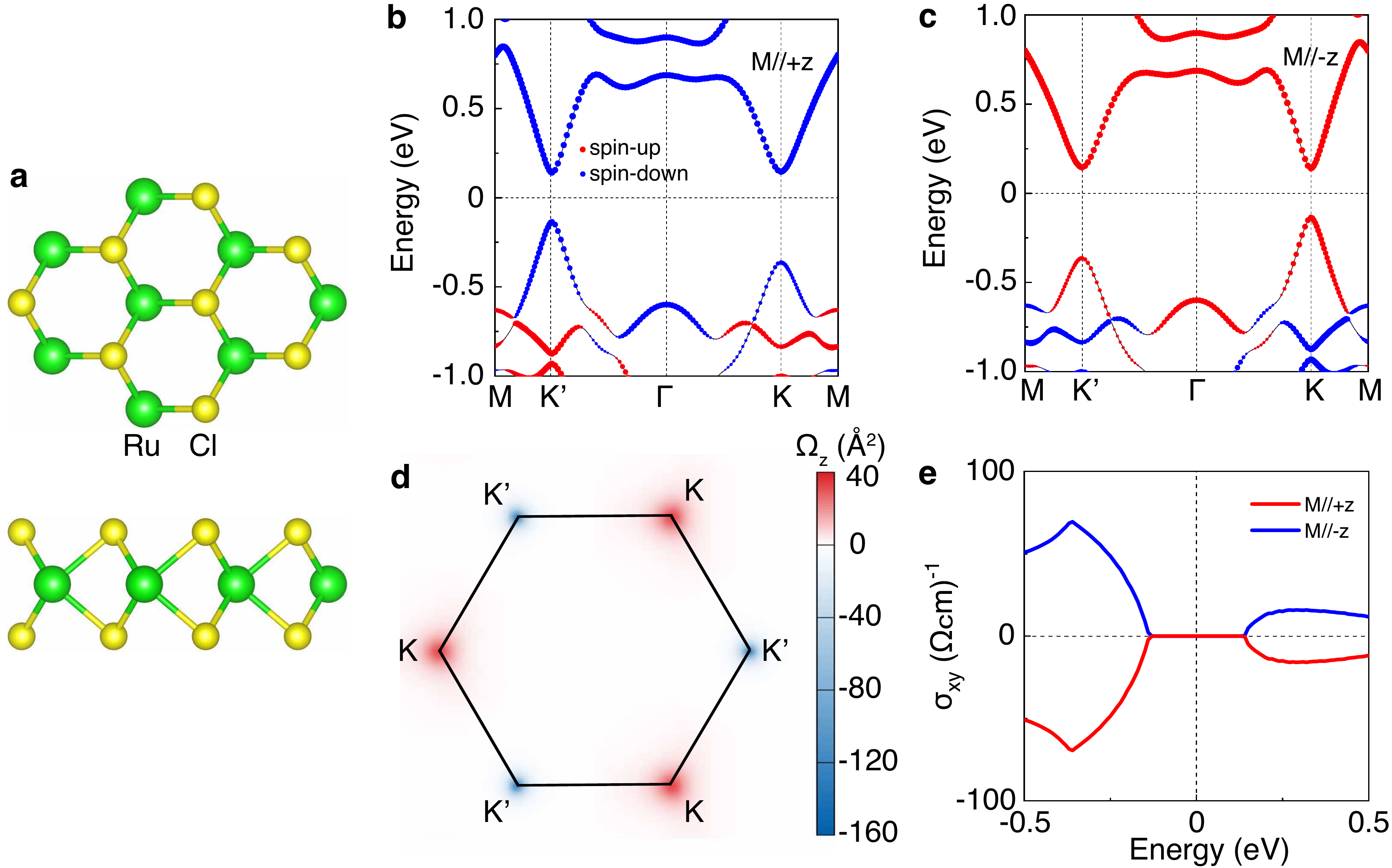}
\caption{\label{fig2} \textbf{Structural, electronic and valleytronic properties of RuCl$_2$ monolayer.} \textbf{a} Top and side view of RuCl$_{2}$ monolayer. \textbf{b}, \textbf{c} Spin-resolved band structures of RuCl$_{2}$ monolayer with magnetization along $+z$ and $-z$ direction. \textbf{d} Berry curvature distribution for the valence bands in the 2D Brillouin zone with $+z$ magnetization. \textbf{e} Anomalous hall conductivity versus the energy eigenvalue for the cases in \textbf{b} and \textbf{c}.} 
\end{figure*}

\section{Results and Discussion}

\subsection{Concept of Half-Valley Ohmic Contact}

When a metal is brought into contact with a semiconductor \cite{tung}, a Schottky or Ohmic MS contact is formed dependent on the relative alignment of the semiconductor band edges and the metal Fermi level $\varepsilon_F$.
In the case of $p$-type MS contact, a Schottky contact is formed if the metal Fermi level is energetically higher than the valence band maximum (VBM) of the semiconductor (Fig. \ref{fig:1}a).
The presence of a potential barrier, known as the Schottky barrier (SB), energetically separate the valence band electronic states from the unoccupied states around the Fermi level in the metal, forming a current-blocking Schottky contact. 
Conversely, if the VBM is higher than the metal Fermi level, the electronic states in semiconductor valence band matches a large abundance of unoccupied states around the metal Fermi level, giving rise to Ohmic contact that allows efficient charge injection across the MS contact (Fig. \ref{fig:1}b). 

In FVSC, the valence band edges at $K$ and $K'$ points of the Brillouine zone are composed of electronic states residing exclusively in $K$ or $K$' valleys, respectively. 
The valence band valley offset of $\Delta_{\text{offset}} \equiv \Delta^{K} - \Delta^{K'}$ where $\Delta^{K(K')}$ is the magnitude of the band edge energy of $K$ or $K'$ valley, respectively, relative to the Fermi level.
Such valley offset creates a wealth of band alignments at a FVSC/metal contact. 
For instance, consider a MS contact composed of FVSC and monolayer graphene which host a matching $K$ and $K'$ valleys to those of the FVSC.
The concept of conventional MS contact can be generalized to such \emph{valleytronic} case, yielding three distinct types contact types  (see Figs. \ref{fig:1}c to \ref{fig:1}d):

\noindent (i) \textbf{Overall Schottky contact}:- The band edges of $K$ and $K'$ valleys of FVSC are both lower than the graphene Fermi level, thus resulting in an \emph{overall Schottky contact} (\ref{fig:1}c) with finite Schottky barrier height (SBH) in both valleys; 

\noindent (ii) \textbf{Overall Ohmic contact}:- The band edges of $K$ and $K'$ valleys of FVSC are both higher than the graphene Fermi level, thus resulting in \emph{overall Ohmic contact} with zero SBH in both valleys (\ref{fig:1}d); 

\noindent (iii) \textbf{Half-valley Ohmic contact}:- Intriguingly, if the band edge energy of only one valley is higher than $\varepsilon_F$ while that of the opposite valley is lower than $\varepsilon_F$, such MS contact forms a peculiar \emph{hybrid} contact type in which one valley forms Ohmic contact with graphene while the opposite valley forms Schottky contact with graphene (\ref{fig:1}e). We regard this contact type as the \emph{half-valley Ohmic contact}, which represents a unique hallmark of valleytronic MS contacts. 

Importantly, current injection is only allowed in one valley of a half-valley Ohmic contact, while being strongly suppressed in the opposite valley (\ref{fig:1}f). 
Valley-contrasting current injection can thus be injected across the FVSC/graphene contact through a half-valley Ohmic FVSC-based MS contact (Fig. \ref{fig:1}g).
As demonstrated below, electrostatic contact gating \cite{yu2009tuning, xiao2020berry} can be utilized to externally tune the band alignment between FVSC band edge energies and graphene Fermi level, thus enabling the FVSC/graphene contact to be switched between various valleytronic contact types.
The valleytronic generalization of MS contact proposed in Fig. \ref{fig:1} thus provide a versatile platform to achieve \emph{contact-limited valley-contrasting carrier injection} - a previously unexplored mechanism to manipulate the valley polarization of the electrical current.

\begin{figure*}
\includegraphics[width=0.958\textwidth]{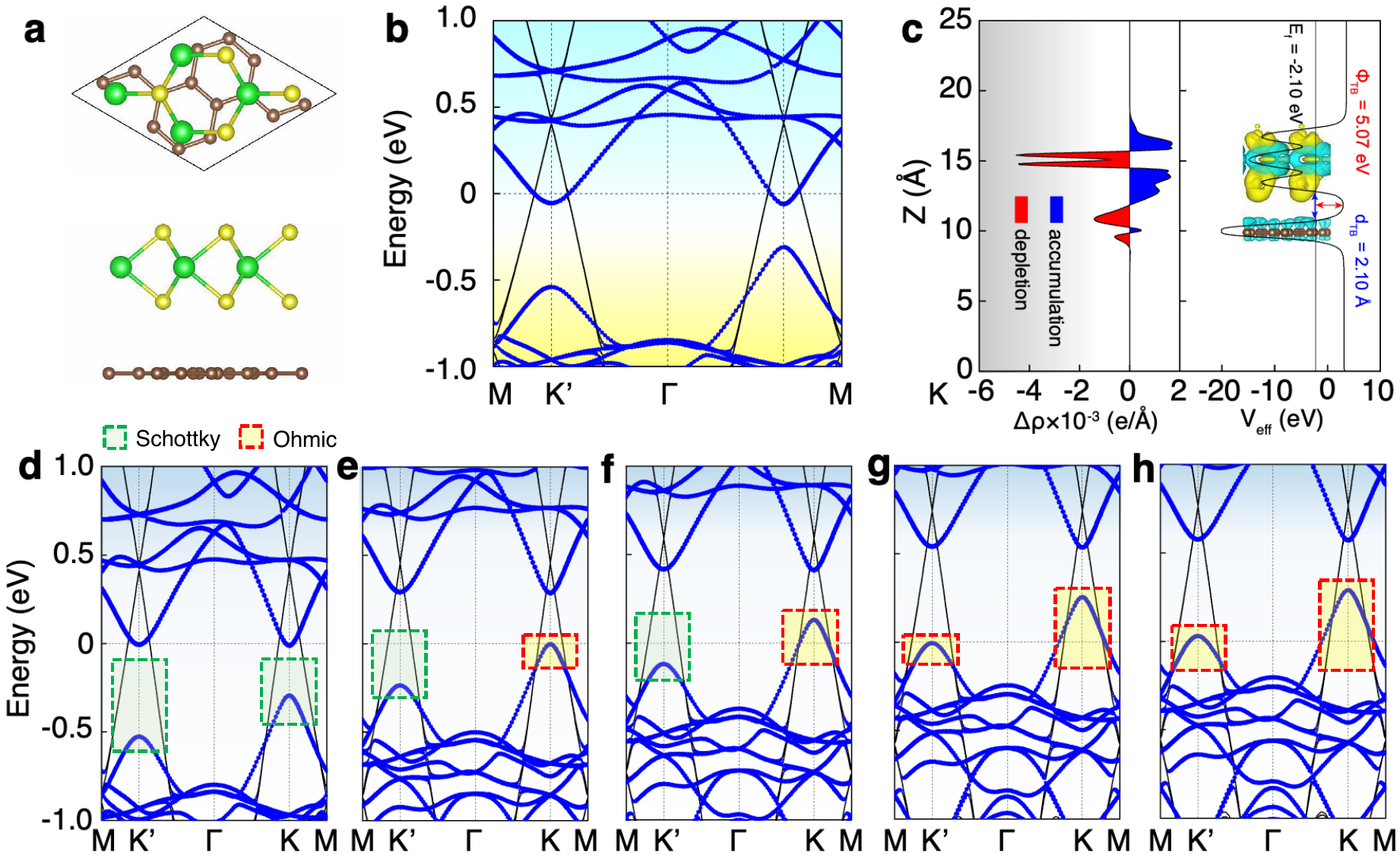}
\caption{\label{fig3}\textbf{Electronic structrues and gate-tunable contact type of RuCl$_2$/graphene contact.} \textbf{a} Top and side view of RuCl$_{2}$/graphene heterostructure. \textbf{b} The band structure of RuCl$_{2}$/graphene with $+z$ magnetization. \textbf{c} The planar differential charge density with isosurface value 0.0001 $e$/{\AA}${\rm ^3}$ and the electrostatic potential across the heterostructure. \textbf{d}-\textbf{h} The band structures of RuCl$_{2}$/graphene with hole doping of 0.1 to 0.5 $e$.}
\end{figure*}

\subsection{Candidate 2D Ferrovalley Semiconductors: RuCl$_2$}

To illustrate the concept of half-valley Ohmic contact, we consider FVSC RuCl$_2$ monolayer (see Fig. \ref{fig2}a for the lattice structures) as the candidate FVSC due to its large spontaneous valley polarization $>200$ meV (see Figs. \ref{fig2}b and \ref{fig2}c for the band structures under opposite magnetization directions) \cite{sheng2022strain,li2022robust}. 
The calculated lattice constants of RuCl$_2$ monolayer is 3.50 {\AA}, which is consistent with previous studies \cite{sheng2022strain,li2022robust}. 
Valley-contrasting transport, such as AVHE \cite{Tong2016}, emerges due to the presence of nontrivial Berry curvature in the phase space, which can be calculated as
\begin{equation}
\Omega_{z}(\boldsymbol{k})=-2 \operatorname{Im} \sum_{n \neq n^{\prime}} f_{n k} \frac{\left\langle n \boldsymbol{k}\left|v_{x}\right| n^{\prime} \boldsymbol{k}\right\rangle\left\langle n^{\prime} \boldsymbol{k}\left|v_{y}\right| n \boldsymbol{k}\right\rangle}{\left(\omega_{n^{\prime}}-\omega_{n}\right)^{2}},
\end{equation}
where ${f}_{nk}$ is the Fermi-Dirac distribution function, ${v}_{x/y}$ are the velocity operators along the $x$/$y$ directions, ${{\varepsilon }_{n}}=\hbar {{\omega }_{n}}$ is the band energy. 
The Berry curvatures of both structures possess unequal absolute values with opposite signs at $K$ and $K'$ points (Fig. \ref{fig2}d) for RuCl$_2$ - a distinctive hallmark of FVSC.

Carriers residing in the two valleys acquire a valley-contrasting anomalous velocity under an external in-plane electric field $\mathbf{E}$,
\begin{equation}
    \boldsymbol{v}=-\frac{e}{\hbar} \mathbf{E} \times \boldsymbol{\Omega}(\mathbf{k})
\end{equation}
in which their opposite transversal motion leads to VAHE -- an exotic transport phenomenon that is practically useful for information processing and storage applications \cite{xiao2020berry} in addition to probing the fundamental Berry phase physics of FVSC. 
When the out-of-plane magnetization is reversed such as via spin-orbit torque effect \cite{li2016spin}, the spontaneous valley polarization (see Figs. \ref{fig2}b and c for the band structures under different magnetization direction) and the anomalous hall conductivity (Fig. \ref{fig2}e) of RuCl$_2$ can thus be fully reversed externally via magnetic tuning knob, such as ferromagnetic substrate \cite{wei2016strong, bora2021magnetic}. 

\subsection{Material Realization of Half-Valley Ohmic Contact: RuCl$_2$/Graphene Heterostructure}

Monolayer RuCl$_2$ pertains the ${\rm D_{3h}}$ point group, which is similar to that of graphene, thus ensuring a good lattice matching with graphene -- a 2D semimetal with similar $K$ and $K'$. 
We construct a valleytroinc MS heterostructures using $2\times 2$ supercells of RuCl$_2$ combined with $\sqrt{7} \times \sqrt{7}$ supercell of graphene (Fig. \ref{fig3}a), which yields a lattice constant of 7.00 {\AA} and a interlayer distance of $3.43$ {\AA} in the fully relaxed heterostructure.
To verify the energetic stability of the heterostructures, the binding energy ${{E}_{b}}$ is calculated as
\begin{eqnarray}
{{E}_{\text{b}}}=\frac{{{E}_{\text{vdWH}}}-{{E}_{\text{FVSC}}}-{{E}_{\text{graphene}}}}{N}
\end{eqnarray}
where $N$ is the atom number of the heterostructure, and ${{E}_\text{vdWH}}$, ${{E}_\text{FVSC}}$ as well as ${{E}_\text{Gr}}$ are the energy of the contact heterostructure, isolated FVSC and isolated graphene, respectively. 
The binding energy of RuCl$_2$/graphene is $-55.1$ meV/supercell, respectively, thus indicating the energetic feasiblity of the heterostructures. 

The band structure of RuCl$_2$/graphene is shown in Fig. \ref{fig3}b. 
A sizable valley polarization of 231 meV is retained the electronic structures of RuCl$_2$ sub-monolayer, which is about $9k_BT$ with reference to room temperature, thus satisfying room temperature operation requirement. 
Importantly, RuCl$_2$/graphene has a `clean' ideal two-valley band structures where the the lower-energy valley is not energetically `buried' below other hole bands. This aspect is in stark contrast to many other \emph{non-ideal} FVSCs, such as NbSe$_2$, in which the hole pocket at the $\Gamma$ point could lead to non-valley signal that masks out the valleytronic signatures from the $K$ and $K'$ valleys (for example, see Fig. S1 in the SI for band structure calculation of NbSe$_2$/graphene heterostructure in which the $\Gamma$-point VBM is undesirably higher than one of the valleys). 
RuCl$_2$/graphene is thus a promising system to realize the half-Ohmic valley contact.

\begin{figure*}
\includegraphics[width=0.985\textwidth]{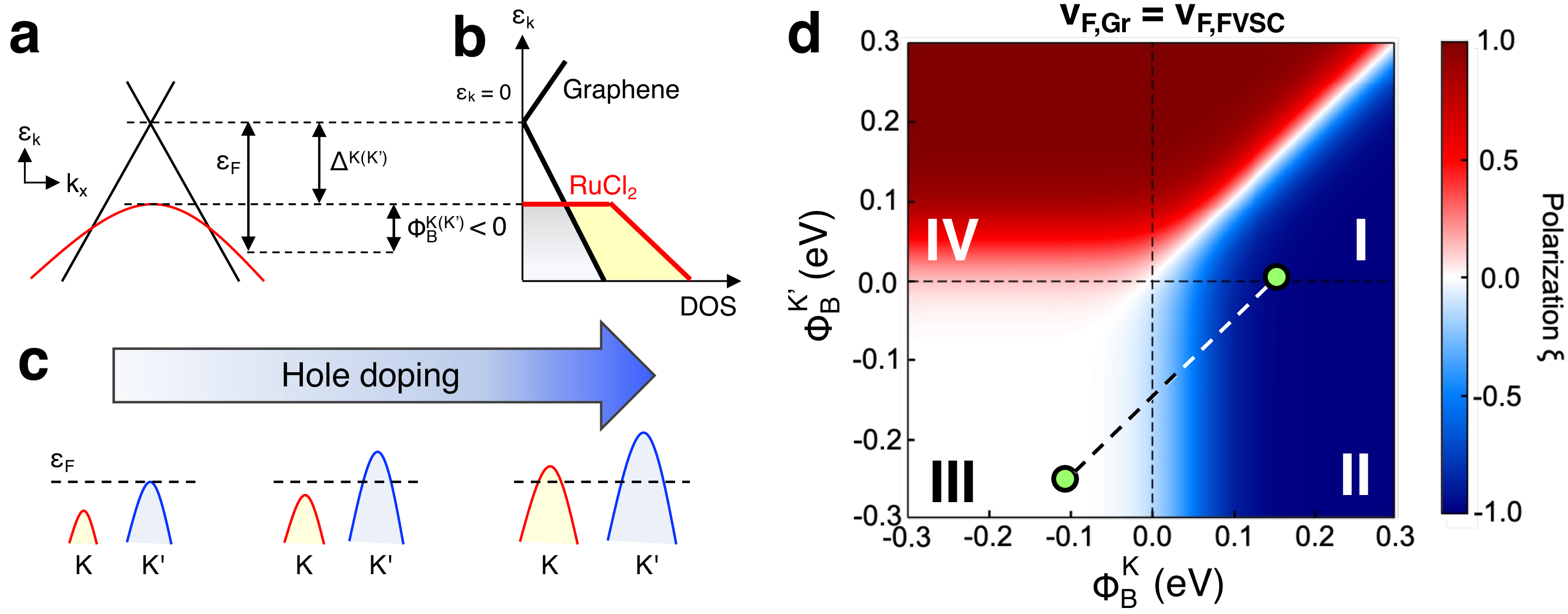}
\caption{\label{fig4}\textbf{Contact-limited valley-contrasting charge injection.} \textbf{a} Band diagram of an Ohmic valley. Here we consider $\Phi_B^{K(K')} < 0$ and $\varepsilon_F < 0$. The zero-energy is taken at the Dirac point of graphene gapless Dirac cone. $\Delta^{K(K')}$ is the valley-dependent energy gap between graphene Dirac point and the VBM.  \textbf{b} Density of states (DOS) corresponding to \textbf{a}. The DOS of RuCl$_2$ is larger than that of graphene due to the `flatter' dispersion of RuCl$_2$. The injection current is limited by graphene due to its smaller DOS. \textbf{c} Schematic drawing of the band structure evolution during hole doping. The contact transforms from half-valley Ohmic to overall Ohmic contacts with increasing hole doping level. \textbf{d} Valley polarization phase diagram. Dashed line denotes the evolution path of $\Phi_B^{K(K')}$ upon electrostatic doping.}
\end{figure*}

RuCl$_2$/graphene heterostructure is intrinsically a non-valleytronic $n$-type Ohmic contact. 
As the work function of RuCl$_2$ is 5.99 eV, which is higher than the work function of graphene \cite{yu2009tuning}, significant electrons transfer from graphene to RuCl$_2$ occurs. This results in $p$-type characteristics of graphene while RuCl$_2$ gains electrons to form $n$-type contact.
The charge transfer process is also revealed in the planar-averaged differential charge density in Fig. \ref{fig3}c in which electron accumulation (depletion) on the RuCl$_2$ (graphene) side is evident. 
A van der Waals gap is present between the two constituent monolayers leads which can be characterized by a plane-averaged tunneling potential barrier \cite{tho2023cataloguing} of height $\Phi_\text{TB} = 5.07$ eV and width $d_\text{TB} = 2.10$ {\AA}.

Although the RuCl$_2$/Gr heterostructure is intrinsically an $n$-type contact, \emph{electrostatic doping} through an external gate voltage can transform the system into $p$-type half-valley Ohmic contact. Hole doping, which can be induced by an external gate voltage \cite{zhang2020electric,chen2020charge, xiao2020berry}, energetically up-shifts the band structure of RuCl$_2$/graphene heterostructure, thus enabling the contact type to be modulated from $n$-type to $p$-type. Such modulation effect is demonstrated in Figs. \ref{fig3}d to \ref{fig3}h where the doping levels are increased from 0.1 to 0.5 hole per supercell. The band structure evolves progressively from $n$-type Ohmic $\to$ $p$-type half Ohmic to $p$-type overall Ohmic contacts. The $p$-type half valley Ohmic contact is achieved at a moderate doping level of 0.3 hole per supercell (Fig. \ref{fig3}f) while $p$-type overall Ohmic contact can be achieved when the doping level exceeds 0.4 hole per supercell. 
Such doping-controlled contact types of RuCl$_2$/graphene heterostructure can be harnessed to achieve gate-tunable valley-polarized current, thus suggesting FVSC/graphene as a versatile building block of valleytronics device applications.

\subsection{Theory of Contact-Limited Valley Current Injection}

We now develop a model of contact-limited current injection across a valleytronic MS contact and demonstrate how such injection process enables valley-polarized current to flow across a FVSC/graphene heterostructure. We consider the case where a bias voltage $V_\text{ds}$ is applied at graphene. The net valence band electrical injection current via a single valley (i.e. $K$ or $K'$ valley) can be expressed as \cite{ang2018universal, Somvanshi2017}
\begin{equation}\label{J_k}
 \mathcal{J}^{K(K')} = \frac{g_s e}{(2\pi)^2 \tau_\text{inj}} \int_\text{B.Z.} \text{d}^2 \mathbf{k}\mathcal{T}\left(\varepsilon_\mathbf{k}, \Phi_B^{K(K')}\right)  \left( f_s - f_m \right)
\end{equation}
where the superscript $K$ ($K'$) denotes $K$ ($K'$) valley, $g_s$ is the spin degeneracy, $\Phi_B^{K(K')}$ is the valley-dependent SBH relative to graphene Fermi level ($\varepsilon_{F}$), $\mathbf{k}$ is the in-plane 2D wave vector of electrons, $\tau_\text{inj}$ is the out-of-plane carrier injection time parameter \cite{sinha2014ideal,javadi2020kinetics} which is material-and device-dependent, $f_m = f( \varepsilon_k +eV_\text{ds})$ and $f_s = f(\varepsilon_k)$ are the carrier distribution function of graphene and FVSC, respectively, with $f(\varepsilon_k)$ being the equilibrium Fermi-Dirac distribution function. 
For over-barrier thermionic emission, the tunneling probability is $\mathcal{T}\left(\varepsilon_\mathbf{k}, \Phi_B^{K(K')}\right) = \Theta\left( \left|\varepsilon_\mathbf{k}\right| - \Phi_B^{K(K')}\right)$ where $\Theta(x)$ is the Heaviside step function. 

Consider isotropic energy dispersion, which is appropriate for graphene and RuCl$_2$, Eq. (\ref{J_k}) becomes
\begin{equation}\label{general_J}
    \mathcal{J}^{K(K')} = \frac{e}{\tau_\text{inj}} \int_{-\infty}^{-(\Phi_B^{K(K')}+\varepsilon_F)} \text{d}\varepsilon_\mathbf{k} \mathcal{D}(\varepsilon_\mathbf{k}) \Delta f(V_\text{ds})
\end{equation}
where we have taken the Dirac point of graphene as the zero-energy, and $\Delta f(V_\text{ds}) \equiv f_s - f_m$. The electronic density of states (DOS) term is defined as $\mathcal{D}(\varepsilon_\mathbf{k}) \equiv \text{min}\left[ \mathcal{D}_\text{Gr}(\varepsilon_\mathbf{k}), \mathcal{D}_\text{FVSC}(\varepsilon_\mathbf{k}) \right]$ since the transport current is limited by the constituent monolayer that has the smaller DOS (Figs. \ref{fig4}a and \ref{fig4}b). The DOS of graphene and FVSC are $\mathcal{D}_\text{Gr}(\varepsilon_\mathbf{k}) = (g_s / 2\pi \hbar^2 v_{F, \text{Gr}}^2) \left|\varepsilon_k\right|$ and $\mathcal{D}_\text{FVSC}(\varepsilon_\mathbf{k}) = \Theta\left( \left|\varepsilon_\mathbf{k}\right| - \Delta/2 \right)(g_s / 2\pi \hbar^2 v_{F, \text{FVSC}}^2) \left|\varepsilon_\mathbf{k}\right| $, respectively. 
The Fermi velocity parameters are extracted from the DFT-calculated band structure as $v_{F, \text{Gr}} \approx 0.87 \times 10^6$ m/s and $v_{F, \text{FVSC}} \approx 0.28 \times 10^6$ m/s for graphene and RuCl$_2$, respectively. 
As $v_{F, \text{Gr}} > v_{F, \text{FVSC}}$, the DOS of RuCl$_2$ is larger than that of graphene due to the `flatter' energy dispersion of the gapped Dirac cone of RuCl$_2$ (Fig. \ref{fig4}b). 
The injection current thus becomes graphene-limited, and can be reduced to
\begin{eqnarray}\label{inj}
  \mathcal{J}^{K(K')} &=& \frac{e}{\tau_\text{inj}}\frac{g_s}{2\pi \hbar^2 v_{F, \text{Gr}}^2 } \int_{-\infty}^{-\Phi_B^{K(K')}} \text{d}\varepsilon_\mathbf{k} \left|\varepsilon_\mathbf{k}\right| \Delta f(V_\text{ds}) 
\end{eqnarray}
Generally, the valley injection current increases with smaller $\Phi_B$. 
As demonstrated in Figs. \ref{fig3}d to \ref{fig3}h, electrostatic hole doping of the graphene/RuCl$_2$ heterostructure results in the up-shifting of the $K$ and $K'$ bands, which transforms the heterostructure from half-valley Ohmic to overall Ohmic contacts (Fig. \ref{fig4}c).
To how $\Phi_B^{K(K')}$ modulation affect the valley-contrasting charge injection in the graphene/FVSC heterostructure, we calculate the valley polarization phase diagram with $\Phi_B^{K(K')}$ varying from -0.3 to 0.3 eV (Fig. \ref{fig4}e), which is defined as 
\begin{equation}
\xi =\frac{\mathcal{J}^{K}(V_\text{ds})-\mathcal{J}^{K'}(V_\text{ds})}{\mathcal{J}^{K}(V_\text{ds})+\mathcal{J}^{K'}(V_\text{ds})}.
\end{equation}
In Fig. \ref{fig4}d, the quadrants II and IV represent the cases of half-valley Ohmic contact (i.e. $\Phi_B^{K}$ and $\Phi_B^{K'}$) have opposite signs) which exhibits large K' and K valley polarization, respectively. 
Quadrant I corresponds to the case of overall Schottky contact (i.e. overall Schottky (i.e. both $\Phi_B^{K}$ and $\Phi_B^{K'}$ are positive-valued). Here the the injection current of the two `Schottky' valleys compete with each others, resulting in a sharp reversal of the valley polarization reverses between the regime of $\Phi_B^{K}>\Phi_B^{K'}$ and $\Phi_B^{K}<\Phi_B^{K'}$. 
Interestingly, the valley polarization can be switched off in the quadrant III as both valleys are Ohmic, i.e. overall Ohmic in which $\Phi_B^{K}<0$ and $\Phi_B^{K'}<0$. 
Considering a half-valley Ohmic contact with representative SBH values of $\Phi_B^{K} \approx 0$ eV and $\Phi_B^{K'} \approx 0.15$ eV and undergoes the SBH modulation via hole doping as indicated in Fig. \ref{fig4}c, and assuming that the energy difference between the $K$ and $K'$ valley VBM remains approximately the same during the SBH modulation, the $\Phi_B^{K}$ and $\Phi_B^{K'}$ evolve along a linear path in the $\Phi_B^{K}$-$\Phi_B^{K'}$ as indicated by the dashed line in Fig. \ref{fig4}e. The valley polarization is switched off during the electrostatic hole doping, thus suggesting that the valley polarization of a valleytronic MS junction can be manipulated by an electrostatic gate.

\begin{figure*}
\includegraphics[width=0.985\textwidth]{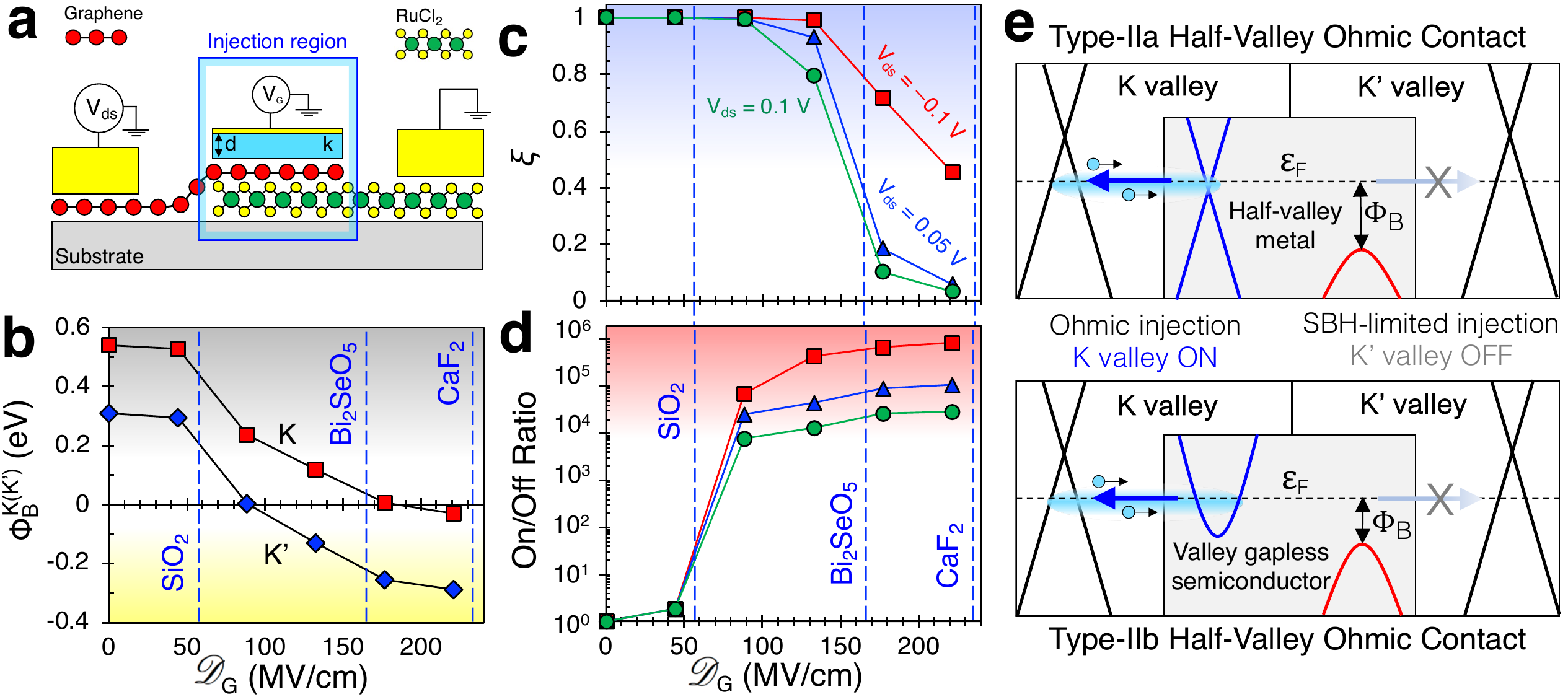}
\caption{\textbf{Concept and performance characteristics of valleytronic barristor.} \label{fig5}\textbf{a} Schematic drawing of a gate-tunable valleytronic barristor. \textbf{b} Schottky barrier height as a function of gate field parameter $\mathscr{E}_G$. The corresponding breakdown condition $k\mathcal{E}_\text{BD}$ for SiO$_2$, HfO$_2$ and TiO$_2$ are indicated by vertical dashed lines. \textbf{c} Valley polarization and \textbf{d} ON/OFF ratio as a function of $\mathscr{E}_G$ with different bias voltage $V_\text{ds}$. \textbf{e} Illustration of half-valley Ohmic contact generalized to the cases of half-valley metal and valley gapless semiconductor.}
\end{figure*}

\subsection{Device Concept and Operation Characteristics of Valleytronic Barristor}

We now demonstrate the valley polarization switching effect based on the DFT simualtion results of graphene/RuCl$_2$.
Figure 5a shows the schematic drawing of a \emph{valleytronic barristor} in which two key functions can be achieved electrostatically via a gate voltage: (i) the valley polarization can be tuned and reserved (ii) the injection current can be switched on and off.
Such gate-tunable operation arises mainly from the electrostatic doping-controlled SBH in the $K$ and $K'$ valleys as described in Figs. \ref{fig3}e to \ref{fig3}h.
It should be noted that the practicality of the electrostatic doping in graphene/RuCl$_2$ is ultimately limited by the gate dielectric breakdown strength. 
To estimate whether the required carrier doping concentration can be achieved within the dielectric breakdown limit of the gate insulators, we employ a planar diode model to obtain the electric displacement field \cite{xia2010graphene}, 
\begin{equation}
    \mathscr{D}_G \equiv \frac{k V_G}{d} = \frac{e n_{\text{ind}}}{\varepsilon_0},
\end{equation}
where $V_G$ is the gate voltage, $d$ is the dielectric thickness, $k$ is the dielectic constant of the dielectric material and $n_{\text{ind}}$ is the electrostatically induced carrier density in the heterostructure. 
Here, $n_{\text{ind}}$ is calculated as the added electron per unit supercell area employed in Figs. \ref{fig3}d to \ref{fig3}h (i.e. 0.1 to 0.5 electron added per unit supercell area). 
The $\mathscr{D}_G$ corresponding to 0.1 to 0.5 $e$ hole doping ranges from about 44 MV/cm to 221 MV/cm (see \textbf{Table S1} for a list of the $\mathscr{E}_G$ values). 
The $\mathscr{D}_G$ is related to the \emph{actual} electric field across the gate dielectric via $\mathcal{E}_\text{diel} = \mathscr{D}_G / k$.
The $\mathscr{D}_G$ can thus be scaled down with high-$k$ gate dielectrics. 
Furthermore, dielectric breakdown limits the maximum hole doping concentration, giving rise to the condition, $\mathscr{D}_G < \mathscr{D}_\text{BD}$ where $\mathscr{D}_\text{BD}= k\mathcal{E}_\text{BD}$ is the electric displacement at the breakdown field of the gate dielectric ($\mathcal{E}_{\text{BD}}$).
Equivalently, $\mathscr{D}_\text{BD}$ corresponds to the maximum carrier density that can be provided by a gate dielectric is $N_\text{max} = (\varepsilon_0 / e) \mathscr{D}_\text{BD}$ \cite{lin2022dielectric}. 
We consider several representative oxide dielectric materials, namely the ${\rm SiO_2}$ ($k = 3.9$, $\mathcal{E} = 10$ MV/cm) widely used in CMOS technology \cite{McPherson2003}, as well as the recently experimentally synthesized single-crystalline layered insulator Bi$_2$SeO$_5$ ($k = 16.5$, $\mathcal{E} = 10$ MV/cm) \cite{zhang2023single} and ultrathin CaF$_2$ ionic crystals ($k = 8.4$, $\mathcal{E} = 27.8$ MV/cm) \cite{wen2020dielectric}. 
Correspondingly, the breakdown electric displacement fields are $\mathscr{D}_\text{BD} = (58.5, 165, 233.5)$ MV/cm for SiO$_2$, Bi$_2$SeO$_5$ and CaF$_2$, respectively. 

The $\varepsilon_C^{K(K')}$ and $\Phi_B^{K(K')}$ are extracted from the DFT-calculated band structures in Figs. \ref{fig3}d to \ref{fig3}h. The simulated transport characteristics are shown in Figs. \ref{fig5}b to \ref{fig5}d. 
The gate dielectric breakdown limits, i.e. $k\mathcal{E}_{BD}$, are marked on Figs. \ref{fig5}b to \ref{fig5}d. 
The $\mathcal{E}_G$ of all hole doping cases can be accommodated by CaF$_2$, thus suggesting a potential experimental feasibility of the electrostatic doping approach in modulating the $K$-and $K'$-valley SBH (Fig. \ref{fig5}b).
Using the contact-limited valley-contrasting current injection model developed above (see Eq. 7), we show in Fig. \ref{fig5}b that the valley polarization can achieved polarization efficiency $>90\%$ and can be tuned towards $0\%$ with higher $\mathscr{E}_G$, thus revealing contact-limited valley-contrasting injection as an effective mechanism to modulate the valley polarization.   
Importantly, Fig. \ref{fig5}d shows that the injection current can exhibit sizable ON/OFF ratio, defined as $\text{ON/OFF ratio} = {J}_\text{total}({\mathcal{E}_{g}})/{J}_\text{total}(0)$ where ${{J}_{\text{total}}}({\mathcal{E}_{g}})$ and ${{J}_\text{total}}(0)$ are the total injection injection with and without hole doping, respectively. Generally, a higher hole doping level (i.e. higher $\mathcal{E}_G$) leads to a more metallic behavior of the contact heterostructure, thus leading to a higher ON/OFF ratio at high $\mathcal{E}_G$. An ON/OFF ratio of $> 10^4$ and above can be achieved.
Figures \ref{fig5}c to \ref{fig5}d suggest that graphene/RuCl$_2$ heterostructure can function as a \emph{valleytronic barristor} in which both the current on/off switching and the valley polarization modulation can be achieved via electrostatic contact gating. 
It should be noted that $\mathscr{D}_G < 150$ MV/cm is sufficient to switch the device into ON state while still maintaining high valley polarization $\xi > 90\%$ (Figs. \ref{fig5}c and \ref{fig5}d). Such a value is well within the breakdown limit of Bi$_2$SeO$_5$ \cite{zhang2023single} and CaF$_2$ \cite{wen2020dielectric} whose integration with 2D semiconductor devices have been recently demonstrated experimentally \cite{illarionov2019ultrathin}. 

Finally, we remark that the half-valley Ohmic contact proposed here can also be achieved using half-valley metal (HVM) \cite{hu2020concepts, PhysRevB.107.054414}, half-valley semiconductor (HVSC) \cite{zhou2021sign} and valley gapless semiconductor (VGSC) \cite{guo2022electric}. Figure \ref{fig5}f schematically illustrate the concept of graphene/HVM contact and graphene/VGSC, which can be regarded as the type-II half-valley Ohmic contacts due to the gapless nature of the band structures. The gapless $K$ valley of the HVM provides barrier-less Ohmic injection into graphene with high injection current, while the injection current through the $K'$ gapped valley is severely suppressed by the Schottky barrier. 
In the VGSC counterpart, the K valley exhibits $n$-type Ohmic contact while the K' valley exhibits $p$-type Schottky contact, which similarly exhibits contact-limited valley contrasting charge injection. 
The possibility of achieving contact-limited valley-contrasting carrier injection using a large variety of valleytronic materials, covering FVSC, HVM and HVSC, suggest the versatility of half-valley Ohmic contact as a potential platform for valleytronics device applications. 

\section{Conclusion}

In summary, we generalized the concept of metal/semiconductor (MS) contact into the valleytronics scenario. We proposed the concept  of half-valley Ohmic contact in graphene/ferrovalley semiconductor (FVSC) heterostructure and proposed its material realization using RuCl$_2$/graphene heterostructure. We developed the theory of contact-limited valley-contrasting charge injection for valleytronic MS contact, and proposed a design of valleytronic barristor in which the valley polarization and the injection current can be switched via a gate voltage under experimentally realizable electrostatic gating conditions. The proposed half-valley Ohmic contact can potentially be realized in valleytronic materials beyond FVSC, such as half-valley metal and valley gapless semiconductor. Our findings revealed the radically different MS contact types in valleytronic materials, and provided a theoretical basis to describe the charge injection process in such contacts. The half-valley Ohmic contact proposed in this work shall provide a useful building block for the development of valleytronic device technology, and shall serve as a harbinger for a new class of valleytronic devices that manipulates valley degree of freedom via contact-limited valley-contrasting charge injection process.

\section*{Methods}

\subsection{Computational Methods}

All density functional theory (DFT) calculations are performed in Vienna Ab-initio Simulation Package (VASP) \cite{Kresse1993Ab,Kresse1996} using projector augmented wave (PAW) method \cite{PAW1994}. The generalized gradient approximation (GGA) with Perdew-Burke-Ernzerhof (PBE) type functional \cite{Perdew1996,kresse1999} is adopted to treat the exchange-correlation interactions between electrons. The GGA+U scheme is adopted to deal with the strong exchange-correlation interactions for Ru-$4d$ orbital. The effective Hubbard $U$ parameter are set as 2.0 eV for ${\rm RuCl_2}$ as implemented in previous work \cite{sheng2022strain}. The kinetic energy is cut off at 500 eV. The 2D reciprocal space is sampled with the Monkhorst-Pack (MP) grid \cite{Monkhorst1976} of $7\times 7\times 1$. The structures are relaxed until the force on each atom converged to 0.001 eV/{\AA} and the energy convergence criterion is $10^{-6}$ eV. A vacuum of 20 {\AA} is applied to eliminate interactions between adjacent layers. The van der Waals interactions is also considered utilizing DFT-D3 approach \cite{Grimme2006}. The Berry curvature and anomalous all conductivity are calculated using WannierTools Package\cite{Wu2018} with tight binding Hamiltonian extracted by WANNIER90 package \cite{Mostofi2008}.

\begin{acknowledgments}
This work is supported by the Singapore Ministry of Education Academic Research Fund Tier 2 (Award No. MOE-T2EP50221-0019) and the SUTD-ZJU IDEA Thematic Research Grant Exploratory Project (SUTD-ZJU (TR) 202203). The computational work for this article was partially performed on resources of the National Supercomputing Centre, Singapore (https://www.nscc.sg).
\end{acknowledgments}

\providecommand{\noopsort}[1]{}\providecommand{\singleletter}[1]{#1}%

\end{document}